# Liquid spreading in trickle-bed reactors: Experiments and numerical simulations using Eulerian-Eulerian two-fluid approach

By

Z. SOLOMENKO[1] ; Y. HAROUN[1] ; M. FOURATI[1] ; F. LARACHI[2] ;

C. BOYER[1] ; F. AUGIER[1]

[1]IFP Energies nouvelles, Rond-point de l'échangeur de Solaize, BP 3, 69360 Solaize, France

[2]Department of Chemical Engineering, Laval University, Québec, QC, Canada, G1V 0A6

**Corresponding Author:** Manel.Fourati@ifpen.fr

Yacine.haroun@ifpen.fr

Tel.: +33 (0)4 37 70 24 97 ; Fax: +33 (0)4 37 70 20 08

**ABSTRACT**

Liquid spreading in gas-liquid concurrent trickle-bed reactors is simulated using an Eulerian two-fluid CFD approach. In order to propose a model that describes exhaustively all interaction forces acting on each fluid phase with an emphasis on dispersion mechanisms, a discussion of closure laws available in the literature is proposed. Liquid dispersion is recognized to result from two main mechanisms: capillary and mechanical (Attou and Ferschneider, 2000; Lappalainen et al., 2009-b). The proposed model is then implemented in two trickle-bed configurations matching with two experimental set ups: In the first configuration, simulations on a 2D axisymmetric geometry are considered and the model is validated upon a new set of experimental data. Overall pressure drop and liquid distribution obtained from γ-ray tomography are provided for different geometrical and operating conditions. In the second configuration, a 3D simulation is considered and the model is compared to experimental liquid flux patterns at the bed outlet. A sensitivity analysis of liquid spreading to bed geometrical characteristics (void-fraction and particles diameter) as well as to gas and liquid flow rates is proposed. The model is shown to achieve very good agreement with experimental data and to predict, accurately, tendencies of liquid spreading for various geometrical bed characteristics and/or phases flow-rates.

# 1. INTRODUCTION

Trickle-bed reactors (TBRs) are widely used in refining industry, particularly in hydro-treatment processes, and continue to mobilize R&D efforts, especially with the growing constraints on sulphur level in diesel and gasoline (Boyer et al., 2005; Martinez et al., 2012). In these reactors catalyst particles are packed to form a fixed bed where, most commonly, liquid and gas reactants flow downwards in cocurrent mode of operation (Wang et al., 2013).

Liquid distribution in trickle bed reactors is a key hydrodynamic criterion for the reactor performance. A poor liquid distribution can basically result in poor utilization or precocious deactivation of the catalyst and may induce local hot spots and low reactor efficiency (Atta et al., 2007; Martinez et al., 2012). It is therefore crucial to investigate liquid spreading in trickle-bed reactors and, ideally, develop predictive models to capture the liquid distribution based on fundamental understanding of hydrodynamic phenomena in such systems. Since pilot-scale experiments are usually time-consuming and costly, their deployment is often restricted to narrow ranges of operating, geometrical and physical conditions (Lappalainen et al., 2011). Numerical modelling has, in response, been indisputably ushering pilot-scale testing as an attractive strategy in studying hydrodynamics in trickle-bed reactors and should help assess proper design of such systems.

Many experimental studies have focused on liquid spreading in trickle-bed reactors. Most of these studies have been carried out on laboratory scale columns equipped with collectors to measure liquid flux patterns at the outlet section of the trickle-bed (Cihla and Schmidt, 1957; Saroha et al., 1998; Li et al., 2000; Marcandelli et al., 2000, Kundu et al., 2001 all cited in Atta et al., 2007). Non-intrusive tomography and video imaging techniques have been also deployed over the last two decades since they allow acquiring quantitative data on liquid distribution at different bed locations (Harter et al., 2001, Boyer and Fanget, 2002; Boyer et al., 2005). Atta et al. (2007) made a literature survey on parameters affecting liquid distribution in trickle-bed reactors. Particles size and shape, particles surface wettability, physicochemical properties of liquid as well as operating gas and liquid flow rates are found to have an impact on liquid spreading. However one should point out a lack of quantitative experimental data that allow proper modelling of liquid spreading for given operational, geometrical and liquid physicochemical properties. In this contribution a new extensive set of experimental data based on γ-ray tomography technique already validated in Boyer and Fanget (2002) and Boyer et al. (2005) is presented. Liquid spreading in a trickle-bed reactor is characterized through liquid hold-up and pressure drop measurements for different particle sizes, and liquid and gas flow rates.

As far as TBRs numerical modelling is concerned, many approaches have been considered hitherto to solve for the two-phase flow in trickle-bed reactors. Wang et al. (2013) made a literature survey of trickle-bed reactors modelling and pointed out two main methods base on Eulerian description, where gas and liquid are treated as interpenetrating continua: volume of fluid method and Euler-Euler method. In the first, a surface tracking technique is used to solve for gas-liquid interface requiring an adapted refined mesh which makes the method applicable for relatively small bed scales (Raynal et al., 2007; Lopes and Quinta-Ferreira, 2010; Haroun et al., 2012; Haroun et al., 2014).

The second, based on an averaging method of local gas and liquid mass and momentum conservation equations is represented through an "effective porous medium" (Wang et al., 2013). Since these models do not simulate directly the flow over the real physical geometry, one should deal with a closure problem where interactions with the solid surface of particles, as well as fluid-fluid interactions should be accounted for through specific closure laws. Wang et al. (2013) summarized the main phenomenological models set up so far to account for these interactions: (i) the relative permeability model (Saez and Carbonell, 1985 ), (ii) the slit model (Holub et al., 1992 ) and (iii) the fundamental force balance model (Attou et al., 1999 in Wang et al., 2013). In order to account for dispersion inherent to the trickle-bed structure, several models have been proposed as well (Grosser et al., 1988; Lappalainen et al., 2009-a&b; Attou and Ferschneider, 2000).

In this contribution, CFD (Computaional Fluid Dynamics) is used to investigate liquid spreading in gas-liquid flow in trickle-bed reactors using Euler-Euler approach. Closure laws as regards fluid-solid interactions, gas-liquid interaction, and exhaustive dispersion mechanisms are discussed based on a literature survey. The retained models have been then implemented as body source terms in momentum conservation equations within ANSYS Fluent 14.5 environment. Simulations were carried out for different particle diameters and void fractions in order to investigate the impact of bed geometry on liquid spreading as well as of gas and liquid flow rates.

Simulation results were then validated upon experimental data of liquid distribution obtained from the literature (Marcandelli et al., 2000 and Boyer et al., 2005) and our own experimental data.

In light of these simulations, a discussion of predominant dispersion mechanisms depending on bed geometrical characteristics as well as operating flow rates is proposed. The ability of the model presented in this paper to assess accurate prediction of liquid spreading in trickle-bed reactors is then highlighted.

## 2. Eulerian-Eulerian two-fluid model

In order to solve for hydrodynamics at the macro-scale of a gas-liquid TBR, a two-fluid approach (Euler-Euler) has been used. Derivation of conservation equations at macro-scale results from averaging pore-scale Navier-Stokes equations over representative volumes that contain significant number of pores but remain small when compared to the trickle-bed size. Derivation of the generic Euler-Euler model equations for porous media using averaging procedure has been described in several works (Whitaker, 1973, 1986, 1999). Moreover, adapted averaging volumes for trickle-bed reactors where liquid flows as a film that is sheared by gas flow have been also proposed by taking into account partial bed wetting (Attou et al., 1999; Iliuta and Larachi, 2005).

Averaging procedure gives rise to several terms containing pore-scale information such as local pressure and velocity deviations. These terms require closure laws in order to solve for the macro-scale problem (Whitaker, 1986; Mewes et al., 1999; Liu, 1999; Lappalainen et al., 2008; Valdes and Parada, 2009). In this work the Euler-Euler model is described and adopted for calculation of gas-liquid co-current flow in a TBR. Closure laws from the literature are discussed and tested over a set of original experimental data as well as data available in literature: Boyer et al. (2005) and Marcandelli et al. (2000).

### 2.1. Mass and momentum conservation equations

Mass and momentum balances are written considering the following average quantities for each phase denoted q (q refers to gas or liquid) over a representative volume V:

- Volume fraction of phase k: $\varepsilon_k = \dfrac{V_k}{V}$ where $V_k$ is the volume of phase k within the representative volume which results in the following geometric relation: $\sum_{k=g,l} \varepsilon_k = \varepsilon$

- Intrinsic average velocity of phase k: $\vec{u}_k = \dfrac{1}{V_k} \int_{V_k} \vec{u}_{pk} dV$ where $\vec{u}_{pk}$ is the interstitial pore-scale velocity of phase k.

- Intrinsic average pressure of phase k: $p_k = \dfrac{1}{V_k} \int_{V_k} p_{pk} dV$ where $p_{pk}$ is the interstitial pore-scale pressure of phase k.

The continuity equation for phase k writes:

$$\frac{\partial}{\partial t}(\varepsilon_k \rho_k) + \vec{\nabla} \cdot (\varepsilon_k \rho_k \vec{u}_k) = 0 \tag{1}$$

The macro-scale momentum balances for gas and liquid phases resulting from averaging procedure over pore-scale Navier-Stokes could be written as follows (Boyer et al., 2005; Lappalainen et al., 2009-a&b; Fourati et al., 2013) :

$$\frac{\partial}{\partial t}(\varepsilon_g \rho_g \vec{u}_g) + \vec{\nabla}.(\varepsilon_g \rho_g \vec{u}_g \vec{u}_g) = -\varepsilon_g \vec{\nabla} p + \vec{\nabla}.(\varepsilon \overline{\overline{\tau}}_g) + \varepsilon_g \rho_g \vec{g} - \vec{F}_{gs} + \vec{F}_{Il} + \vec{F}_{D,g} \qquad (2)$$

$$\frac{\partial}{\partial t}(\varepsilon_l \rho_l \vec{u}_l) + \vec{\nabla}.(\varepsilon_l \rho_l \vec{u}_l \vec{u}_l) = -\varepsilon_l \vec{\nabla} p + \vec{\nabla}.(\varepsilon \overline{\overline{\tau}}_l) + \varepsilon_l \rho_l \vec{g} - \vec{F}_{ls} + \vec{F}_{Ig} + \vec{F}_{D,l} \qquad (3)$$

In equations (2) and (3), the explicit pressure gradient term is based on gas phase pressure ($p=p_g$). $\vec{\nabla}.(\varepsilon \overline{\overline{\tau}}_k)$ (k=g,l) is the divergence of the average stress tensor. This term depends on the intrinsic average velocity of the considered phase and writes as follows - assuming a constant phase viscosity over the averaging volume -: $\vec{\nabla}.(\varepsilon \overline{\overline{\tau}}_k) = \mu_k \nabla^2 \vec{u}_k$. Since the gradient of average velocity is negligible comparing to this of local velocity fluctuations within a pore (Whitaker, 1999), one could neglect the latter term.

Moreover, $\vec{F}_{gs}$ and $\vec{F}_{ls}$ refer, respectively, to gas-solid and liquid-solid interaction forces due to porous resistances (Fourati et al., 2013). Gas-liquid interactions are accounted for using the shear term $\vec{F}_{Ik}$ analogous to drag force in dispersed gas-liquid flows.

$\vec{F}_{D,g}$ and $\vec{F}_{D,l}$ are dispersion forces in the porous medium that apply, respectively, to gas and liquid, and include capillary as well as mechanical effects (Lappalainen, 2009 a,b).

Finally, the momentum balance at gas-liquid interface writes: $\vec{F}_{Il} + \vec{F}_{Ig} = \vec{0}$. (4)

## 2.2. Closure laws

### 2.2.1. Interaction forces

**a/ Porous resistances and gas-liquid interaction**

Proposals to model gas-solid and liquid-solid interactions in porous media are mostly inspired from Darcy-Forchheimer formalism (Whitaker, 1996) that allows modeling the resistance induced by the medium geometry in a given phase. With respect to this formalism, gas-solid and liquid-solid interactions include viscous and inertial forces that could be written for the three flow directions owing to viscous and inertial resistance tensors : $\bar{\bar{D}}_k$ and $\bar{\bar{C}}_k$ (Fourati et al., 2013).

$$\vec{F}_{ks} = -\left( \mu_k \bar{\bar{D}}_k \vec{u}_k + \frac{1}{2} \rho_k \|\vec{u}_k\| \bar{\bar{C}}_k \vec{u}_k \right) \quad (5)$$

where k accounts for gas or liquid phases.

In order to model resistance tensors, several suggestions have been made. Ergun-like models as well as relative permeability models are widely used to account for porous resistances in the particular case of trickle-bed reactors (Saez and carbonell, 1985; Attou et al., 1999; Boyer et al., 2005; Iliuta et al., 2000). Ergun-like models are mostly inspired from the phenomenological one-phase Ergun model assuming a different hydraulic diameter for gas and liquid when these are in contact. On that basis, Holub et al. (1992) suggested viscous and inertial resistances that depend on liquid saturation, bed geometrical characteristics (void and specific surface area) as well as Ergun constants. However the resulting macroscopic model does not take into account gas-liquid interactions that become important when gas and/or liquid inertia increase. Attou et al. (1999) built a phenomenological macroscopic hydrodynamic model in trickle-bed reactors based on the balance of forces exerted on both phases at a particle scale. Within this formalism, the suggested porous resistance that applies to liquid is weighted by the medium tortuosity, itself estimated as the reciprocal of the liquid saturation. Starting from this same macroscopic model, Boyer et al. (2007) conjectured that tortuosity in liquid phase depends, via liquid film curvature (and thus surface tension), on gas inertia (Narasimhan et al., 2002; Kundu et al., 2003; Boyer et al., 2007). Based on experimental data for aqueous and organic liquids, the authors suggested tortuosity to write as $\alpha_L^n$ where n values are respectively -0.53 and -0.02.

Unlike in Holub et al. (1992) model, Boyer et al. (2007) model includes gas-liquid interaction force developed previously in Attou et al. (1999) based on a mechanistic approach. The resulting macroscopic model allows prediction of liquid saturations and overall pressure drop with relative errors as low as 8% and 16%, respectively. Lopes and Quinta-Ferreira (2009) used this phenomenological model as well in order to provide closure laws for TBR simulations. Authors reported relative errors on simulated values of liquid hold-up and pressure drop as low as 2% provided an adapted mesh density is considered. We have thus retained Boyer et al. (2007) model to account for porous resistances as well as gas-liquid interactions in the present work.

By viewing trickle-bed reactors as isotropic media, porous resistances described in equation (5) on both phases write as follows (Boyer et al., 2007):

$$\vec{F}_{ks} = -K_{ks}\vec{u}_k \tag{6}$$

In equation (6), fluid-solid momentum exchange coefficients $K_{ks}$ are expressed as follows:

$$K_{gs} = E_1 \mu_g \frac{(1-\varepsilon_g)^2}{\varepsilon_g d_p^2}\left(\frac{1-\varepsilon}{1-\varepsilon_g}\right)^{2/3} + E_2 \rho_g \frac{1-\varepsilon_g}{d_p}\left(\frac{1-\varepsilon}{1-\varepsilon_g}\right)^{1/3}\|\vec{u}_g\| \tag{7}$$

$$K_{ls} = \alpha_l^n \left( E_1 \mu_l \frac{(1-\varepsilon)^2}{\varepsilon_l d_p^2} + E_2 \rho_l \frac{1-\varepsilon}{d_p}\|\vec{u}_l\| \right) \tag{8}$$

Gas-liquid interaction force contains viscous and inertial contributions as well and writes as follows:

$$\vec{F}_{Ig} = K_{gl}\left(\vec{u}_g - \vec{u}_l\right) \tag{9}$$

Where $K_{gl}$ is a gas-liquid interaction coefficient that writes as follows:

$$K_{gl} = E_1 \mu_g \frac{(1-\varepsilon_g)^2}{\varepsilon_g d_p^2}\left(\frac{1-\varepsilon}{1-\varepsilon_g}\right)^{2/3} + E_2 \rho_g \frac{1-\varepsilon_g}{d_p}\left(\frac{1-\varepsilon}{1-\varepsilon_g}\right)^{1/3}\|\vec{u}_g - \vec{u}_l\| \tag{10}$$

In closure laws expressed in equations (7), (8) and (10), viscous and inertial Ergun constants $E_1$ and $E_2$ are taken equal to 150 and 1.75. These have been validated upon pressure drop measurements in single-phase upward liquid flow on 5 different liquid-solid systems (Boyer et al., 2007).

**b/ Capillary dispersion**

The two-fluid model allows solving for a single averaged pressure field that equals gas pressure. Therefore a capillary pressure model is needed to account for the pressure jump, $p_g - p_l$ across gas-liquid interface due to local film curvature. The gradient of capillary pressure induces capillary dispersion that applies for the liquid by chosing gas-phase pressure as the dependent pressure variable in the model formulation (Lappalainen et al., 2008; Fourati et al., 2013).

Several authors have proposed capillary pressure models (Leverett, 1941; Grosser et al., 1988; Attou and Ferschneider, 2000; Wilhite et al., 2005; Lappalainen et al., 2009-b). Grosser et al. (1988) proposed empirical model to relate capillary pressure to liquid saturation based on Leverett (1941) experiments. The model has been questioned since then and was shown to behave well in regimes of moderate liquid saturation but fails in reproducing pendular regimes that correspond to low liquid saturations (Lappalainen et al., 2009b). One could also point out the tendency of the model to predict non-physical subsequent capillary pressure when bed void fraction tends to 1 making it inapplicable for large particle diameters. Lappalainen et al. (2009-b) reminded that the relevance of Leverett's function used in Grosser et al. (1988) model is not known very well in the case of relatively large particles used in trickle-bed reactors.

Lappalainen et al. (2009-b) proposed a capillary pressure empirical model based on local geometrical characteristics of gas-liquid interface curvature. Two capillary regimes –funicular and pendular- have been considered based on the shape of the gas-liquid interface. Curvature is described using polynomial functions that involve experimentally fitted coefficients over experiments of Dodds and Srivastava (2006). The resulting model, based on three polynomial functions, gives good tendencies of capillary pressure in pendular regime (sharp rise of capillary pressure when liquid saturation decreases) but still with important discrepancies in that particular regime.

Attou and Ferschneider (2000) proposed a capillary pressure model based on Young-Laplace equation. Two main curvatures of the gas-liquid interface have been thus considered. The first involves a characteristic diameter $d_1$ defined as the equivalent diameter of a solid particle covered by a liquid film. The second uses an equivalent hydraulic diameter $d_2$ of an interstitial channel approximated by the minimum diameter of three contacting spheres. The authors proposed expressions for gas-liquid interface curvatures that involve liquid saturation, bed void-fraction and particle diameters based on geometrical considerations. In order to account for the decrease of gas-liquid interface curvature at elevated operating pressures consistent with increasing gas density, Attou and Ferschneider (2000) proposed an ad hoc correction factor, $F(\rho_G/\rho_L)$, on capillary pressure. Such factor has been proposed based on Wammes et al. (1990, 1991) experimental results on transition from trickling to pulse flow regime. Attou and Ferschneider (2000) model has been validated upon a large range of experimental data regarding the onset of transition from trickling to pulse regime using different particle diameters, particle shapes and pressure. As compared to the model of Lappalainen et al. (2009b), this model does not match with static capillary pressure data from literature based on capillary suction method (Dodds and Srivastava, 2006) for very low liquid saturations. However, the model predicts very well capillary pressure in regimes of moderate liquid saturation in which trickle-bed reactors commonly operate. For the reasons cited above and since Attou and Ferschneider (2000) model does not include empirical variables, it has been retained in this work to account for capillary pressure in Euler-Euler simulations in a gas-liquid trickle-bed reactor. The latter model has also the advantage of being relatively simple to implement in a CFD simulation.

Therefore, the capillary pressure writes:

$$p_c = p_g - p_l = 2\sigma\left(\frac{1}{d_p} + \frac{1}{d_{\min}}\right) \times F\left(\frac{\rho_g}{\rho_l}\right) \times \left(\frac{1-\varepsilon}{1-\varepsilon+\varepsilon\alpha_l}\right)^{1/3} \qquad (11)$$

Where $F\left(\frac{\rho_g}{\rho_l}\right) = \left(1 + 88.1\frac{\rho_g}{\rho_l}\right)$ accounts for the effect of gas density on capillary pressure.

From equation (11) one could express capillary saturation gradient that gives rise to capillary liquid dispersion as follows:

$$\vec{\nabla} p_c = -1 \times \frac{2\sigma}{3} \times \left(\frac{\varepsilon}{1-\varepsilon}\right) \times \left(\frac{1}{d_p} + \frac{1}{d_{\min}}\right) \times F\left(\frac{\rho_g}{\rho_l}\right) \times \left(\frac{1-\varepsilon}{1-\varepsilon+\varepsilon\alpha_l}\right)^{4/3} \times \vec{\nabla}\alpha_l \qquad (12)$$

In equation (12) the characteristic diameter $d_{\min}$ is defined as the minimum equivalent diameter of the area delineated between three contacting spheres and is expressed as follows:

$$d_{\min} = \left(\frac{\sqrt{3}}{2} - \frac{1}{2}\right)^{\frac{1}{2}} \times d_p$$

**c/ Mechanical dispersion**

Mechanical dispersion is a phenomenon that occurs in packed bed reactors in general due to the particular geometry of the porous medium in which the fluids are allowed to evolve. Lappalainen et al. (2009b) suggested that mechanical dispersion is essential to predict liquid spreading and has increasing effect as particle size goes up. Fourati et al. (2013) discussed the mechanistic origin as well as the theoretical foundation of mechanical dispersion forces in packed beds. These authors interpreted dispersion as due to local spatial fluctuations of pore-scale liquid velocity at pore scale in the packed bed. It is indeed the macroscopic velocity deviation from the volume-averaged velocity which *per se* causes dispersion. From a theoretical point of view the authors pointed out that the term expressing the divergence of local velocity deviations correlation tensor (Whitaker et al., 1973) could be understood as a mechanical dispersion term. A few studies have proposed macroscopic generic forms to model mechanical dispersion (Mewes et al., 1999; Liu and Long, 2000) but explicit closure laws remain scarce. Lappalainen et al. (2009b) suggested a mechanical dispersion force analog that accounts for diffusion of passive tracers. The authors proposed cast mechanical dispersion forces in both gas and liquid in terms of the so-called spread factor ($S_m$) and specific drift velocities. Mechanical dispersion forces in liquid and gas write, respectively, as follows (Lappalainen et al., 2009b, 2011, Fourati et al., 2013):

$$\vec{F}_{D,mech,l} = \alpha_l K_{ls} \vec{u}_{D,l} + \varepsilon\, K_{gl} \left( \vec{u}_{D,l} - \vec{u}_{D,g} \right) \tag{13}$$

$$\vec{F}_{D,mech,g} = \alpha_g K_{gs} \vec{u}_{D,g} + \varepsilon\, K_{gl} \left( \vec{u}_{D,g} - \vec{u}_{D,l} \right) \tag{14}$$

In these equations, $K_{ks}$ (k=g,l) and $K_{gl}$ are, respectively, momentum exchange coefficients and gas-liquid interaction coefficients previously expressed in equations (7), (8) and (10).

Drift velocities for liquid and gas are functions of fluid saturation gradient and write as follows (Lappalainen et al., 2009b):

$$\vec{u}_{D,k} = -\frac{S_m}{\alpha_k} \left( \|\vec{u}_k\| \vec{\nabla} \alpha_k - (\vec{u}_k \cdot \vec{\nabla} \alpha_k) \frac{\vec{u}_k}{\|\vec{u}_k\|} \right) \ (k=l,g) \tag{15}$$

Spread factor $S_m$ is usually determined from liquid spreading experiments starting from a single liquid feeding point. Comprehensive representations to predict spread factors for different particle sizes and shapes are quite scarce (Lappalainen et al., 2009b). These latter compared several literature approaches to experimental data and showed that Baldi and Specchia (1976) representation was the least unsatisfactory. The same authors conducted tracer dispersion experiments for particles of different sizes and shapes (beads, Berl saddles and Raschig rings) for the water-air system. They extended the correlation of Onda et al. (1973) to account for particle shapes. For spherical particles, the resulting correlation is:

$$S_m = 0.015 \times \sqrt{d_p} \tag{16}$$

This correlation allows satisfactory agreement with Baldi and Specchia (1976) experimental data as well as data from Porter et al. (1968) and Cihla and Smith (1958).

Lappalainen et al. (2009 b) also hypothesized that dispersion acts on both liquid and gas. In trickle-bed reactors where liquid flows mainly as films on the packing surface, mechanical dispersion impacts, in the first place, mostly the liquid phase as pointed out by Fourati et al. (2013) and confirmed via numerical simulations. In the present simulations, exhaustive modeling of dispersion in both gas and liquid is considered.

## 3. Experimental data base

In this section two main sets of experiments are presented. Experimental data will be then compared to simulations at the same geometrical and operating conditions.

### 3.1. Liquid spreading from a source point : new database based on experiments of Boyer et al. (2005)

Boyer et al. (2005) measured liquid distribution in a 0.4 m diameter and 1.8 m high packed-bed filled with spherical glass beads using γ-ray tomography technique. They considered an air-water system and a diameter of particles of 1.99 mm. Liquid was fed from a central nozzle located at top of the column at a volumetric flow rate of 128 l/h. Gas was injected from a lateral nozzle for two volumetric flow rates: 45 and 90 $m^3$/h. The experimental apparatus used by authors is described on Figure 1.

γ-ray tomography was used to acquire liquid saturation profiles at different column sections located respectively at 60, 280 and 780 mm from liquid inlet (Figure 1). This tomography technique as well as the reconstruction algorithm used to acquire liquid saturation maps have been extensively described and validated in Boyer and Fanget (2002). The uncertainty on liquid saturation has been shown to be less than ± 3%.

Moreover, Boyer et al. (2005) determined a bed void-fraction profile for 1.99 mm particles along the packed bed by measuring height and mass of successive loaded bed slices of 20 cm height. The resulting bed void-fraction profile is given by the following polynomial function (where z is the distance from bed top as shown in Figure 1):

$$\varepsilon(z) = 0.421 - 0.124z + 0.047z^2 \tag{17}$$

In addition to these measurements, a new set of data based on Boyer et al. (2005) experiments is added in terms of particles diameter, liquid and gas flow rates (see Table 1). Measurements were actually carried out using the same column geometry, γ-ray tomography technique, liquid inlet configuration and locations of liquid saturation profiles as described in Boyer et al. (2005), see also Figure 1. Pressure drop across the trickle-bed was also measured.

### 3.1 3.2 Liquid spreading from off-center inlet distribution : experiments of Marcandelli et al. (2000)

Marcandelli et al. (2000) investigated liquid distribution in a 0.3 diameter and 1.3 high column packed with 2 mm glass beads and using air-water system. These authors considered different liquid and gas feeding configurations and measured liquid flow rate using 9 collectors of constant surface at the bottom of the column. In this study, we chose only off-center liquid feeding configuration where gas was fed through four 25 mm ID chimneys while liquid is injected through two 2.5mm ID orifices as illustrated in Figure 2 in the cocurrent mode. Unlike the symmetrical point injection described in section 3.1, this liquid inlet geometry allows judging the capability of Euler-Euler model prediction under 3D maldistribution configurations.

In order to quantify liquid distribution at the bed outlet, Marcandelli et al. (2000) defined a "maldistribution index" ranging from 0 to 1 and calculated as follows:

$$M_f = \sqrt{\frac{1}{N(N-1)}\sum_i \left(\frac{Q_i - Q_{mean}}{Q_{mean}}\right)^2} \qquad (18)$$

In equation (18), N accounts for the number of collectros (9 in this case), $Q_i$ for the liquid flow rate through a given collector referenced by i and $Q_{mean}$ for the mean flow rate defined as an arithmetic average of flow rates (equal area sampling domains) collected at bottom of the packed bed.

Experimental data have been provided in terms of liquid distribution profiles by means of % of total liquid flow rate in each collector as well as maldistribution index for given liquid and gas superficial velocities. These authors indicated relative errors on measured liquid flow rates in collectors ranging from 8 to 10%.

## 4. Simulation Results

Simulations have been carried out within Ansys Fluent 14.5 CFD environment. The basic Euler-Euler formalism available within this software has been extended with User defined Functions to account for the specific closure laws described in section 2.

An unsteady-state solver with implicit temporal discretization scheme of order 2 has been used. Upwind scheme of order 2 has been considered for spatial discretization.

### 4.1. Liquid spreading from a source point

#### 4.1.1. Simulation conditions

In order to simulate liquid spreading experiments described in section 3.1, a two-dimensional computational grid has been used considering an axisymmetric condition on the column axis. The longitudinal bed extension has been reduced to 1 m in order to reduce calculation time. 2 mm x 4 mm rectangular grid cells have been considered.

Velocity-inlet conditions have been used at the domain inlet and a pressure-outlet condition has been specified at the domain exit.

Most of simulations have been carried out using constant bed void-fraction: 37% in the case of 1.99 mm particles and 41% in the case of 6 mm particles. Case 1 (Table 1) has been simulated using both constant bed void-fraction and axial void-fraction profile described in section 3.1. However, in all simulation cases, bed void fraction is assumed constant in the radial direction conceiving that a ratio $d_p/D_c$ is 20 is large enough to ignore radial void effects (Boyer et al., 2005; Herskowitz and Smith, 1978).

### 4.1.2. Results and discussion

In Figure 3, liquid saturation fields as well as liquid jet contours in simulation and experimental cases described in Table 1 are reported at different longitudinal distances from liquid inlet: respectively, at 60, 280 and 780 mm as defined in Boyer et al. (2005).

Liquid jet contours are defined as the liquid jet width at a given longitudinal coordinate denoted z of the trickle-bed.

Experimentally, liquid jet width is determined via the reconstruction algorithm described in Boyer et al. (2005). The liquid jet extent is supposed to be reached when a sharp density gradient $\frac{\partial \rho_k}{\partial r}$ is detected in the radial direction, indicating the interface between a liquid-rich and a gas-rich region. Numerically, determination of the extent of liquid jet for given conditions requires a criterion on liquid saturation or derivative thereof. In the first case, a threshold of saturation at the liquid jet limit is to be considered making liquid jet contours dependent on that specific value. In this study, liquid jet width is defined as the radial coordinate r at which a discontinuity is observed on liquid saturation gradient in the radial direction: $\frac{\partial \alpha_l}{\partial r}$. This criterion is deemed more objective than the first one.

Different criteria for determination of liquid jet extent may induce artificial discrepancies between experimental and numerical data in terms of liquid spreading for the same conditions. Thus, one should keep this in mind when comparing liquid jet widths obtained experimentally and numerically.

On Figure 4, liquid saturation radial derivatives $\frac{\partial \alpha_l}{\partial r}$ obtained from simulations are shown as function of radial coordinate r at different longitudinal positions in the packed-bed. Figure 4 shows clearly sharp increase in liquid saturation gradient that indicates liquid jet extent. In Table 2, the values of liquid jet width, denoted $R_{jet}$, are given for experimental and inherent simulation cases described in Table 1.

### 4.1.2.1 Effect of bed void fraction on liquid jet spreading

Case 1 (Table 1) was simulated considering constant bed void-fraction of 0.37 then considering longitudinal void-fraction profile determined in Boyer et al. (2005). Simulation results in both cases are presented in Table 3.

Implementation of longitudinal void-fraction profile obtained experimentally allows slight improvement in pressure drop prediction along the bed. In this case, constant relative error of 30% is obtained on pressure drop in two longitudinal zones (Table 3). As far as liquid jet width is concerned, implementing a variable void-fraction profile does not influence simulation results. Consequently, one could consider constant bed void fraction amply sufficient to predict liquid spreading in this bed geometry without altering the quality of model prediction.

**4.1.2.2 Effect of particles diameter on liquid jet spreading**

In order to determine the effect of particle diameter on liquid spreading, one should compare respectively case 1 and case 4, case 2 and case 5, case 3 and case 6 (Figure 3 and Table 2). Experimental data in terms of $R_{jet}$ (Table 2) show more important liquid jet spreading for particles of diameter 6 mm as compared to 1.99 mm ones at intermediate and lower longitudinal positions (z = 280 mm and z = 780 mm). However, for upper longitudinal position (z = 60 mm), $R_{jet}$ remains globally slimmer for 6 mm particles than for 1.99 mm particles.

As far as simulations are concerned, globally less subsequent liquid jet spreading across the bed is predicted for 6 mm particles (Figure 3 and Table 2). Lappalainen et al. (2009b) reported equally a slight decrease in liquid spreading when replacing 6.35 mm diameter particles with 9.53 mm diameter ones, based on radial liquid flux profiles from Herskowitz and Smith (1978).

Liquid spreading shortage for coarser particle observed in this study is essentially due to the decrease of capillary dispersion forces with particle diameter. On the other hand, mechanical dispersion forces become more important as particle diameter increases as learnt from equation 16. However such increase is insufficient to compensate for the loss due to capillary forces thus generating less jet spreading for 6 mm particles.

Evolution of the radial component of mechanical and capillary dispersion forces are shown on Figure 6 for $d_p$=1.99 mm and $d_p$= 6 mm. Figure 6 shows that liquid spreading is governed by capillary dispersion forces for small particles and mechanical dispersion forces for coarser particles as has been underlined in Lappalainen et al. (2009b) as well.

Moreover Figure 3 shows very good agreement between experimental and predicted profiles of liquid jet for larger particle diameter. Relatively higher relative errors are observed for smaller particle. This implies, in all likelihood, that mechanical dispersion, though more accurately described in the present model is not sufficient considering that current capillary dispersion force tends to inflate liquid spreading for particles of small diameter.

**4.1.2.3 Effect of liquid flow rate on liquid jet spreading**

In order to investigate liquid flow rate effect on liquid spreading in trickle-bed reactor, one should compare respectively results inherent in case 1 and case 3 ($d_p$= 1.99 mm) then case 4 and case 6 ($d_p$= 6 mm) described in Table 1.

Experimental data given in Figure 3 and Table 2 show that an increase in liquid flow rate leads liquid jet to widen at upper and intermediate elevations of the trickle-bed (~35% wider) then to tighten for lower longitudinal positions (~12% tighter). Such observations are persistent for both particle sizes examined in this work.

Simulations predict basically liquid jet widening near the trickle-bed top (~18% wider) when liquid flow rate increases but liquid jet extent remains almost unchangeable at deeper downstream bed locations. Increase of liquid jet width near bed inlet for higher liquid flow rates is explained by important liquid velocity at the bed entrance in one point injection conditions. This results in higher mechanical dispersion force (Equations 13, 14 and 15) and consequently wider liquid jet at top of trickle-bed.

**4.1.2.4 Effect of gas flow rate on liquid jet spreading**

In order to investigate liquid flow rate effect on liquid spreading, one should compare respectively results inherent in case 1 and case 2 ($d_p$= 1.99 mm) then case 4 and case 5 ($d_p$= 6 mm) described in Table 1.

According to experimental data given in Figure 3 and Table 2, increasing gas flow rate does not influence liquid spreading near the bed inlet but causes jet confinement for lower longitudinal positions (~15% tighter liquid jet). This is basically observed irrespective of the particle sizes examined in this work.

As far as simulation results are concerned, no gas effect on liquid spreading is observed. As a matter of fact, the criterion used in this work to determine liquid jet extent (section 4.1.2) does not allow reproducing jet confinement at the bottom of the bed. In order to make visible such phenomenon, one should rather focus on liquid saturation profiles.

Figure 7 presents the evolution of constant liquid saturation lines obtained from simulations in cases 1 and 2. Liquid jet confinement is then well reproduced by simulations since constant liquid saturation lines are translated towards smaller radial coordinates when gas flow rate increases. Based on Figure 7, liquid jet width reduction is only ~7%. More important liquid jet confinement is obtained in experiments (Table 2).

A few experimental studies have dealt with gas effect on liquid spreading (Moller et al., 1996; Saroha et al., 1998; Kundu et al., 2001). Gas flow rate is recognized to have a moderate impact on liquid distribution in trickle-bed reactors as compared to liquid flow rate. Globally, gas flow is shown to improve liquid distribution, for a given liquid low rate, and reduce flow in the near-wall region when liquid is injected from uniform distributor. This is due to increase of pressure drop that leads to better radial liquid spreading when gas flow rate is increased (Moller et al., 1996; Kundu et al., 2001). Llamas et al. (2009) investigated gas flow influence on liquid distribution in trickle-bed reactors using two experimental techniques: wire-mesh tomography to measure local liquid saturation and collectors to measure liquid flux at the bed outlet. They pointed out a detrimental impact of gas flow-rate on liquid distribution, in terms of saturation profiles on a bed section, for a non-uniform incident liquid distribution. Moreover, these authors warned that liquid flux patterns measured from a liquid collecting technique may lead to appreciation error as regards liquid distribution because of the inability of this technique to capture maldistributions in terms of liquid saturation. Results obtained in this work as regards gas effect on liquid spreading seem consistent with Llamas and al. (2009) observations.

## 4.2. Liquid spreading from off-center inlet distribution : experiments of Marcandelli et al. (2000)

In this section, experiments of Marcandelli et al. (2000) have been considered. Contrarily to liquid feeding configuration from a central nozzle, these experiments allow test the present developed model in off-center feeding conditions resulting in 3D non-axisymmetric flow.

Simulations have been carried out considering 3D grid that represents half of the column described in section 3.1. The simulation domain has been chosen so that symmetry of the flow is fulfilled. Gas and liquid, flowing in the co-current mode, have been introduced in the domain through velocity inlet conditions and gauge pressure has been specified at the trickle-bed outlet. Superficial velocities of gas and liquid used for simulation are, respectively, 3E-03 m/s and 51E-03 m/s (Marcandelli et al., 2000). Trickle-bed void fraction was considered constant and set equal to 39 % (Llamas, 2009). Relatively small time steps, varying between 1E-05 s and 1E-02 s, were required for this simulation.

Figure 8 shows liquid saturation contours in the simulated trickle-bed in the particular configuration of two liquid injection points (Figure 2-a) considered in Marcandelli et al. (2000) experiments. One could notice liquid spreading along the column. Dissymmetry of liquid patterns is kept but still attenuated at lower positions of the trickle-bed.

In order to characterize liquid distribution at the bed outlet, ratio of liquid flow rate in each virtual collector (Figure 2-b) to total liquid flow rate is calculated. Figure 9 shows very good agreement between experimental and numerical data bearing in mind relative errors on measured liquid flow rates that range between 8 and 10%. Moreover simulation predicts a global maldistribution factor (Equation 18) of 27% which is quite close to this reported by Marcandelli et al. (2000) which value is 23%.

Atta et al. (2007) attempted to simulate Marcandelli et al. (2000) experiments as well. The model proposed by the authors predicts relatively well liquid distribution in homogeneous inlet conditions but failed in predicting the flow in the case of decentred liquid injection. This is probably due to the absence of dispersion forces in the k-fluid model considered by the authors. In fact, simulations carried out in this study without taking into account capillary and mechanical dispersion forces showed negligible spreading of liquid in the trickle-bed.

## 5. CONCLUSION

In this work, an Euler-Euler model is proposed and validated to simulate the hydrodynamics in trickle-bed reactors in terms of liquid spreading. Closure laws are discussed in order to account for fluid-solid interactions, fluid-fluid interactions as well as dispersion forces induced by the medium geometry and phasic inertia. The model includes exhaustive description of dispersion mechanisms that contribute to liquid dispersion, namely, capillary (Attou and Ferschneider, 2000) and mechanical (lappalainen et al., 2009-b). In order to validate the model, simulations have been conducted in experimental conditions described in the studies of Boyer et al. (2005) and Marcandelli et al. (2000).

In the first configuration, liquid is injected in the trickle bed from a centered single feeding point located at top of the bed. The second configuration corresponds, however, to an off-center injection resulting in truly 3D flow patterns.

Calculations in the first feeding conditions are compared to a new set of experimental data based on γ-ray tomography validated in Boyer et al. (2005) regarding different bed geometries and operating conditions. Simulations show satisfactory agreement with experimental data in terms of pressure drop and deliver an acceptable magnitude of liquid jet extent at different bed positions. A sensitivity analysis to bed geometry in terms of void fraction and particle sizes has been carried out. It has been shown that mechanical dispersion is predominant over capillary dispersion the larger the particle diameter. Accuracy of mechanical dispersion model has been pointed out based on excellent agreement with experimental data. The impact of gas and liquid flow rates on liquid spreading has been also investigated. Liquid flow rate is shown to promote liquid spreading especially near the bed inlet with much less extent farther downwards. Conversely, gas flow rate increase is shown to bring about only a moderate liquid jet confinement across the entire trickle-bed domain.

Finally, simulations of off- center liquid injection conditions described in Marcandelli et al. (2000) were possible owing to 3D model simulations allowing accurate prediction of liquid flow patterns in this case.

In order to enhance the model prediction for different geometrical and operating conditions, extensive experimental data are still required in order to confirm liquid spreading tendencies for significant range of particles diameter and phases flow rates. Development of adapted capillary dispersion models for trickle-bed reactors at different operating regimes is still required.

Provided it is used in conjunction with pilot-scale experimental studies, CFD is shown to be a powerful and complementary asset to predict flow-structure hydrodynamic phenomena in trickle-bed reactors. Indeed, the Eulerian-Eulerian two-fluid model can be used to investigate phases distribution in TBRs at industrial column scale. For instance, this allows to evaluate the quality of distribution for different particle grading sizes and to study the evolution of mal-distribution. Moreover hydrodynamic simulations using two-fluid model can be coupled with chemical species transport and chemical reaction. In the case of hydro-treatment processes for instance, this approach allows to relate the quality of distribution and the catalytic performance and thus better define design criteria for the distribution technologies.

# References

Atta A., Roy S., Nigam K.D.P., 2007. Investigation of liquid maldistribution in trickle-bed reactors using porous media concept in CFD. Chem. Eng. Sci, vol. 62, pp. 7033-7044.

Attou A., Boyer C., Ferschneider G., 1999. Modelling of the hydrodynamics of the cocurrent gas-liquid trickle flow through a trickle-bed reactor. Chem. Eng. Sci, vol. 54, pp. 785-802.

Attou A., Ferschneider G., 2000. A two-fluid hydrodynamic model for the transition between trickle an pulse flow in a cocurrent gas-liquid packed-bed reactor. Chem. Eng. Sci, vol. 55, pp. 491-511.

Baldi, G., Specchia, V., 1976. Distribution and radial spread of liquid in packed towers with two-phase cocurrent flow: effect of packing shape and size. Quaderni dell'Ingegnere Chimico Italiano, vol. 12, pp. 107–111.

Boyer C., Fanget B., 2002. Measurement of liquid flow distribution in trickle bed reactor of large diameter with a new gamma-ray tomographic system. Chem. Eng. Sci, vol. 57, pp. 1079-1089.

Boyer C., Koudil A., Chen P., Dudukovic MP., 2005. Study of liquid spreading from a point source in a trickle-bed via gamma-ray tomography and CFD simulation. Chem. Eng. Sci, vol. 60, pp. 6279-6288.

Boyer C., Volpi C., Ferschneider G., 2007. Hydrodynamics of trickle bed reactors at high pressure: Two-phase flow model for pressure drop and liquid holdup, formulation and experimental validation. Chem. Eng. Sci, vol. 62, pp. 7026-7032.

Cihla Z., Schmidt O., 1958. Studies of the behaviour of liquids when freely trickling over the packing of cylindrical tower. Coll Czech Chem Commun, vol. 23, pp. 569-577.

Dodds, J.A., Srivastava, P., 2006. Capillary pressure curves of sphere packings: Correlation of experimental results and comparison with predictions from a network model of pore space. Particle & Particle Systems Characterization, vol. 23, pp. 29−39.

Fourati M., 2012. Dispersion de liquide dans les écoulements gaz-liquide à contre-courant dans les colonnes à garnissages : étude expérimentale et modélisation numérique. Doctoral Thesis, INPT, Toulouse, France.

Fourati, M., Roig V., Raynal L., 2013. Liquid dispersion in packed columns: Experiments and numerical modeling. Chem. Eng. Sci, vol. 100, pp. 266-278.

Grosser K., Carbonell R.G., Sundaresan S., 1988. Oneset of pulsing in two-phase flow cocurrent downflow through a packed bed. A.I.Ch.E. Journal, vol. 34, pp. 1850-1860.

Gunjal P.R., Kashid M.N., Ranade V.V., Chaudhari R.V., 2005. Hydrodynamics of trickle-bed reactors : experiments and CFD modeling. Industrial & Engineering Chemistry Research, vol. 44, pp. 6278-6294.

Haroun Y., Raynal L., legendre D., 2012. Mass transfer and liquid hold-up determination in structured packing by CFD. Chem.Eng.Sci, vol. 75, pp. 342-348.

Haroun Y., raynal L., Alix P., 2014. Prediction of effective area and liquid hold-up instructured packings by CFD. Chem. Eng. Res. Des.

Harter I., Boyer C., Raynal L., Ferschneider G. and Gauthier T., 2001. Flow Distribution Studies applied to Deep Hydro-Desulfurization. Ind. Eng. Chem. Res., vol.40, pp. 5262-5267.

Holub R. A., Dudukovic M.P., Ramachandran P.A., 1992. A phenomenological model for pressure drop, liquid holdup, and flow regime transition in gas-liquid trickle flow. Chem. Eng. Sci, vol. 47, pp. 2343-2348.

Herskowitz M., Smith J.M., 1978. Liquid distribution in trickle-bed reactors. A.I.Ch.E Journal, vol. 24, pp. 439-450.

Iliuta I. and Larachi F., 2005. Modelling the Hydrodynamics of Gas-Liquid Packed Beds via Slit Models: A Review. International Journal of Chemical Reactor Engineering, vol. 3, Review 4.

Jiang Y., Khadilkar M.R., Al-Dahhan M.H., Dudukovic M.P., 2002. CFD of multiphase flow in packed-bed reactors: I. k-fluid modeling issues. A.I.Ch.E Journal, vol. 48, pp. 701-715.

Kundu, A., Saroha, A.K., Nigam, K.D.P., 2001. Liquid distribution studies in trickle bed reactors. Chem. Eng. Sci, vol. 56, pp. 5963–5967.

Kundu, A., Nigam, K.D.P., Verma, R.P., 2003. Catalyst wetting characteristics in trickle bed reactors. A.I.Ch.E. Journal, vol. 49 (9), pp. 2253–2263.

Lappalainen K., Alopaeus V., Manninen M., Aittamaa J., 2008. Improved hydrodynamic model for wetting efficiency, pressure drop, and liquid holdup in trickle bed reactors. Industrial & Engineering Chemistry Research, vol. 47, pp. 8436-8444.

Lappalainen K., Manninen M., Alopeus V., Aittamaa J., Dodds J., 2009-a. An analytical model for capillary pressure-saturation relation for gas-liquid system in a packed-bed of spherical particles. Transport in Porous Media, vol. 77, pp. 17-40.

Lappalainen K., Manninen M., Alopaeus V., 2009-b. CFD modeling of radial spreading of flow in trickle-bed reactors due to mechanical and capillary dispersion. Chem. Eng. Sci, vol. 64, pp. 207-218.

Lappalainen K., 2009-c. Modelling gas liquid flow in trickle-bed reactors. Doctoral thesis, Helsinki University of Technology.

Lappalainen K., Gorshkova E., Manninen M., Alopeus V., 2011. Characteristics of liquid and tracer dispersion in trickle-bed reactors: effect on CFD modeling and experimental analyses. Computers and Chemical Engineering, vol. 35, pp. 41-49.

Leverett M.C., 1941. Capillary behaviour in porous solid. AIME Transactions, vol. 142, pp. 152-168.

Li, M., Iida, N., Yasuda, K., Bando, Y., Nakamura, M., 2000. Effect of orientation of packing structure on liquid flow distribution in trickle bed. Journal of Chemical Engineering of Japan, vol. 33, pp. 811–814.

Liu S., 1999. A continuum approach to multiphase flows in porous media. Journal of Porous Media, vol. 2, pp. 295-308.

Liu S., Long J., 2000. Gas-liquid countercurrent flows through packed towers. Journal of Porous Media, vol. 3, pp. 99-113.

Llamas J-D., Lesage F., Wild G., 2009. Influence of Gas Flow Rate on Liquid Distribution in Trickle-Beds Using Perforated Plates as Liquid Distributors. Ind. Eng. Chem. Res., vol. 48, pp. 7–11.

Llamas J-D., 2009, Etude expérimentale de la maldistribution des fluides dans un réacteur à lit fixe en écoulement co-courant descendant de gaz et de liquide, Doctoral thesis, Institut National Polytechnique de Lorraine.

Lopes R.J.G, Quinta-Ferreira R.M, 2009. CFD modelling of multiphase flow distribution in trickle-beds. Chemical Engineering Journal, vol. 147, pp. 342-355.

Lopes, R. J. G. and R. M. Quinta-Ferreira, ''Assessment of CFD-VOF Method for Trickle-Bed Reactor Modelling in the Catalytic Wet Oxidation of Phenolic Wastewaters,'' Ind. Eng. Chem. Res. 49(6), 2638–2648 (2010).

Marcandelli C., Lamine A.S., Bernard J.R., Wild G., 2000. Liquid distribution in trickle-bedreactor. Oil & Gas Science and Technology, vol. 55, 407-415.

Martinez M., Pallares J., Lopez J., Lopez A., Albertos F., Garcia MA., Cuesta I., Grau FX., 2012. Numerical simulation of the liquid distribution in a trickle-bed reactor. Chem. Eng. Sci, vol. 76, pp. 49-57.

Mewes D., Loser T., Millies M., 1999. Modelling of two-phase flow in packings and monoliths. Chem. Eng. Sci, vol. 54, pp. 4729-4747.

Moller L-B., Halken C., Hansen J-A., Bartholdy J., 1996. Liquid and gas distribution in trickle-bed reactors. Ind. Eng. Chem. Res., vol. 35, pp. 926-930.

Narasimhan, C.S.L., Verma, R.P., Kundu, A., Nigam, K.D.P., 2002. Modeling hydrodynamics of trickle-bed reactors at high pressure. A.I.Ch.E. Journal, vol. 48, pp. 2459–2474.

Onda K., Takeuchi H., Maeda Y., Takeuchi N., 1973. Liquid distribution in packed column. Chem. Eng. Sci, vol. 28, pp. 1677-1683.

Porter, K.E., Barnett, V.D., Templeman, J.J., 1968. Liquid flow in packed columns. Part II: the spread of liquid over random packings. Transactions of the Institution of Chemical Engineers, vol. 46, pp. T74–T85.

Raynal, L., Royon-Lebeaud, A., 2007. A multi-scale approach for CFD calculations of gas–liquid flow within large size columnequipped with structured packing. Chem. Eng. Sci., vol. 62, pp. 7196–7204.

Saez A.E., Carbonell R.G., 1985. Hydrodynamic parameters for gas-liquid cocurrent flow in packed beds. A.I.Ch.E Journal, vol. 31, pp. 52-62.

Saroha, A.K., Nigam, K.D.P., Saxena, A.K., Kapoor, V.K., 1998. Liquid distribution in trickle-bed reactors. A.I.Ch.E. Journal, vol. 44, pp. 2044–2052.

Valdes-Parada F.J., Ochoa-Tapia J.A., Alvarez-Ramirez J., 2009. Validity of the permeability Carman-Kozeny equation: A volume averaging approach, Physica A, vol. 388, pp. 789-798.

Wammes, J.A., Westerterp, K.R., 1990. The influence of the reactor pressure on the hydrodynamics in a concurrent gas-liquid trickle-bed reactor. Chem. Eng. Sci, vol. 37, pp. 1849–1962.

Wammes, J.A., Westerterp, K.R., 1991. Hydrodynamics in a concurrent gas–liquid trickle-bed at elevated pressures. A.I.Ch.E. Journal, vol. 45, pp. 2247–2254.

Wang Y., Jinwen C., Larachi F., 2013. Modelling and simulation of trickle-bed reactors using computational fluid dynamics: a state-of-the-art review. The Canadian Journal of Chemical Engineering, vol. 91, pp. 136-180.

Wilhite B.A., Blackwell B., Kacmar J., Varma A., McCready M.J., 2005. Origins of Pulsing Regime in Cocurrent Packed-Bed Flows. Ind. Eng. Chem. Res, vol. 44, pp. 6056-6066.

Whitaker S., 1973. The transport equations for multi-phase systems. Chem. Eng. Sci, vol. 28, pp. 139-147.

Whitaker S., 1986. Flows in porous media II: The governing equations for immiscible two-phase flow. Transport in Porous Media, vol. 1, pp. 105-125.

Whitaker S., 1999. The Method of Volume Averaging, Kluwer Academic Publishers, Dordrecht.

# Nomenclature

Greek Symbols

| | |
|---|---|
| $\alpha_k$ | Saturation of phase k [-] |
| $\varepsilon_k$ | Volume fraction of phase k [-] |
| $\varepsilon$ | Bed void-fraction [-] |
| $\mu_k$ | Dynamic viscosity of phase k, [kg/m.s]. |
| $\rho_k$ | Mass density of phase k, [kg/m$^3$]. |
| $\sigma$ | Surface tension, [N/m]. |

Latin Symbols

| | |
|---|---|
| $\overline{\overline{C}}_k$ | Inertial and inertial resistance tensors, [m$^{-1}$] |
| $\overline{\overline{D}}_k$ | Viscous and inertial resistance tensors, [m$^{-2}$] |
| $d_p$ | Particles diameter, [m]. |
| $E_1, E_2$ | Ergun constants, [-]. |
| $\vec{F}_{gs}, \vec{F}_{ls}$ | Gas-solid and liquid-solid interaction force, [Nm$^{-3}$]. |
| $\vec{F}_{ll}$ | Gas-liquid interaction force, [Nm$^{-3}$]. |
| $\vec{F}_{D,k}$ | Total dispersion force exerted on phase k, [Nm$^{-3}$]. |
| $\vec{g}$ | Gravity vector, [m/s$^2$]. |
| $M_f$ | Maldistribution index, [-]. |
| N | Number of liquid collectors, [-]. |
| $p$ | Pressure, [Pa]. |
| $p_c$ | Capillary pressure, [Pa]. |
| $Q_i$ | Liquid flow rate on collector of index I, [m$^3$/s]. |
| $Q_{mean}$ | Mean liquid flow rate, [m$^3$/s]. |
| $Q_G$ | Gas flow, [m$^3$/s]. |
| $Q_L$ | Liquid, [m$^3$/s]. |
| $S_m$ | Spread factor, [m] |
| $\vec{u}_k$ | Velocity vector of phase k, [m/s]. |
| $\vec{u}_{D,k}$ | Drift velocity vector of phase k, [m/s]. |
| $V_k$ | Volume of phase k [m$^3$] |
| $V_{SL}$ | Superficial liquid velocity, [m/s] |
| $V_{SG}$ | Superficial gas velocity, [m/s] |
| V | Elementary volume [m$^3$] |

Sub/superscripts

| | |
|---|---|
| $k=g, l$ | Gas and Liquid. |
| $i$ | Collector index |
| r,z | Radial and longitudinal coordinates, [m] |

## List of Figures

## List of Tables

# Tables

**Table 1 : operational and geometrical conditions of liquid spreading experiments**

| Case | 1 | 2 | 3 | 4 | 5 | 6 |
|---|---|---|---|---|---|---|
| | Boyer et al. (2005) | Boyer et al. (2005) | New | New | New | New |
| Particle diameter (mm) | 1.99 | 1.99 | 1.99 | 6 | 6 | 6 |
| Water flow rate (L/h) | 128 | 128 | 453 | 128 | 128 | 453 |
| Air flow rate (m3/h) | 45 | 90 | 45 | 45 | 90 | 45 |

**Table 2 : Liquid jet width R$_{jet}$ at different longitudinal positions for simulation cases described in Table 1. Comparison between experimental data and simulations results**

|  |  | R$_{jet}$ at 60 mm from injection | R$_{jet}$ at 280 mm from injection | R$_{jet}$ at 780 mm from injection |
|---|---|---|---|---|
| **Case 1** Q$_L$= 128 l/h Q$_G$=45 m³/h d$_p$=1.99 mm | Experiment | 52 mm | 72 mm | 94 mm |
|  | Simulation | 64 mm | 100 mm | 124 mm |
|  | Relative Error (%) | 23 | 39 | 32 |
| **Case 2** Q$_L$= 128 l/h Q$_G$=90 m³/h d$_p$=1.99 mm | Experiment | 52 mm | 70 mm | 62 mm |
|  | Simulation | 64 mm | 102 mm | 126 mm |
|  | Relative Error (%) | 23 | 46 | 103 |
| **Case 3** Q$_L$= 453 l/h Q$_G$=45 m³/h d$_p$=1.99 mm | Experiment | 69 mm | - | 80 mm |
|  | Simulation | 76 mm | 102 mm | 124 mm |
|  | Relative Error (%) | 10 | - | 55 |
| **Case 4** Q$_L$= 128 l/h Q$_G$=45 m³/h d$_p$=6 mm | Experiment | 44 mm | 77 mm | 140 mm |
|  | Simulation | 44 mm | 76 mm | 106 mm |
|  | Relative Error (%) | 0 | -1 | -24 |
| **Case 5** Q$_L$= 128 l/h Q$_G$=90 m³/h d$_p$=6 mm | Experiment | 52 mm | 90 mm | 120 mm |
|  | Simulation | 46 mm | 78 mm | 110 mm |
|  | Relative Error (%) | -12 | -13 | -8 |
| **Case 6** Q$_L$= 453 l/h Q$_G$=45 m³/h d$_p$=6 mm | Experiment | 62 mm | 100 mm | 125 mm |
|  | Simulation | 58 mm | 86 mm | 114 mm |
|  | Relative Error (%) | -6 | -14 | -9 |

**Table 3 : Simulation and experimental results (Boyer et al., 2005) in case 1 for constant and variable bed void-fraction**

|  |  | $\varepsilon$= 0.37 | | $\varepsilon$= f (z) | |
|---|---|---|---|---|---|
|  | experiments | simulation | Relative error (%) | Simulation | Relative error (%) |
| R$_{jet}$ (mm) : 60 mm from injection | 52 | 65 | 25 | 62 | 19 |
| R$_{jet}$ (mm) : 280 mm from injection | 72 | 102 | 42 | 101 | 40 |
| R$_{jet}$ (mm) : 780 mm from injection | 94 | 129 | 37 | 128 | 36 |
| Δp/Δz (Pa/m) ( z in [0 500 mm]) | 984 | 867 | -12 | 679 | -31 |
| Δp/Δz (Pa/m) ( z in [500 1000 mm] | 1736 | 1031 | -41 | 1232 | -29 |

# Figures

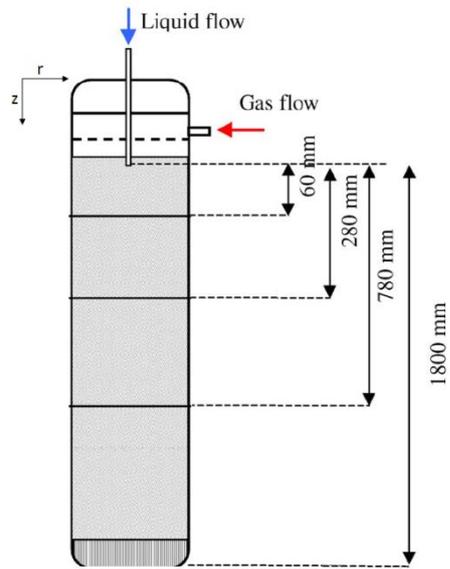

**Figure 1 : Boyer et al. (2005) experimental apparatus**

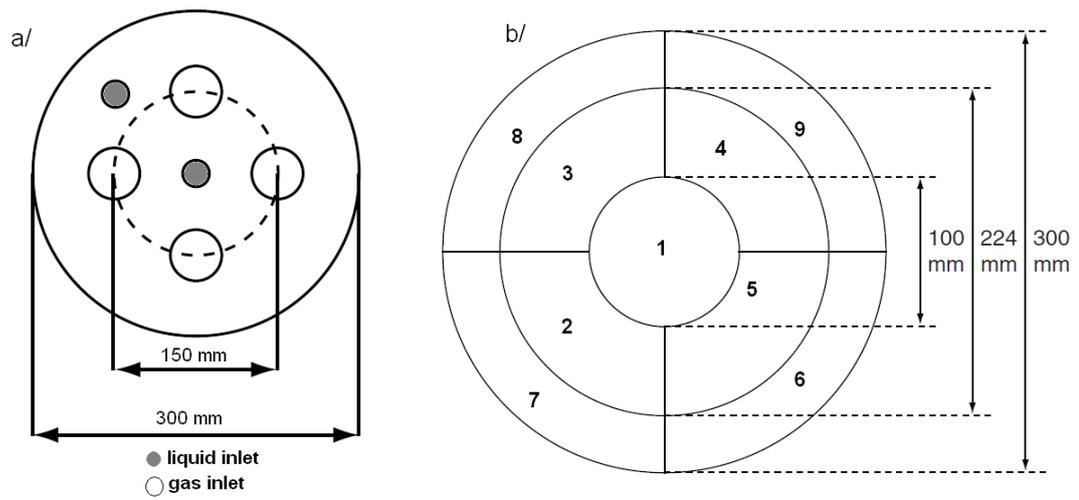

**Figure 2 : Marcandelli et al. (2000) experiments configuration. a/ feeding geometry at top of the bed ; b/ collectors geometry at bottom of the bed**

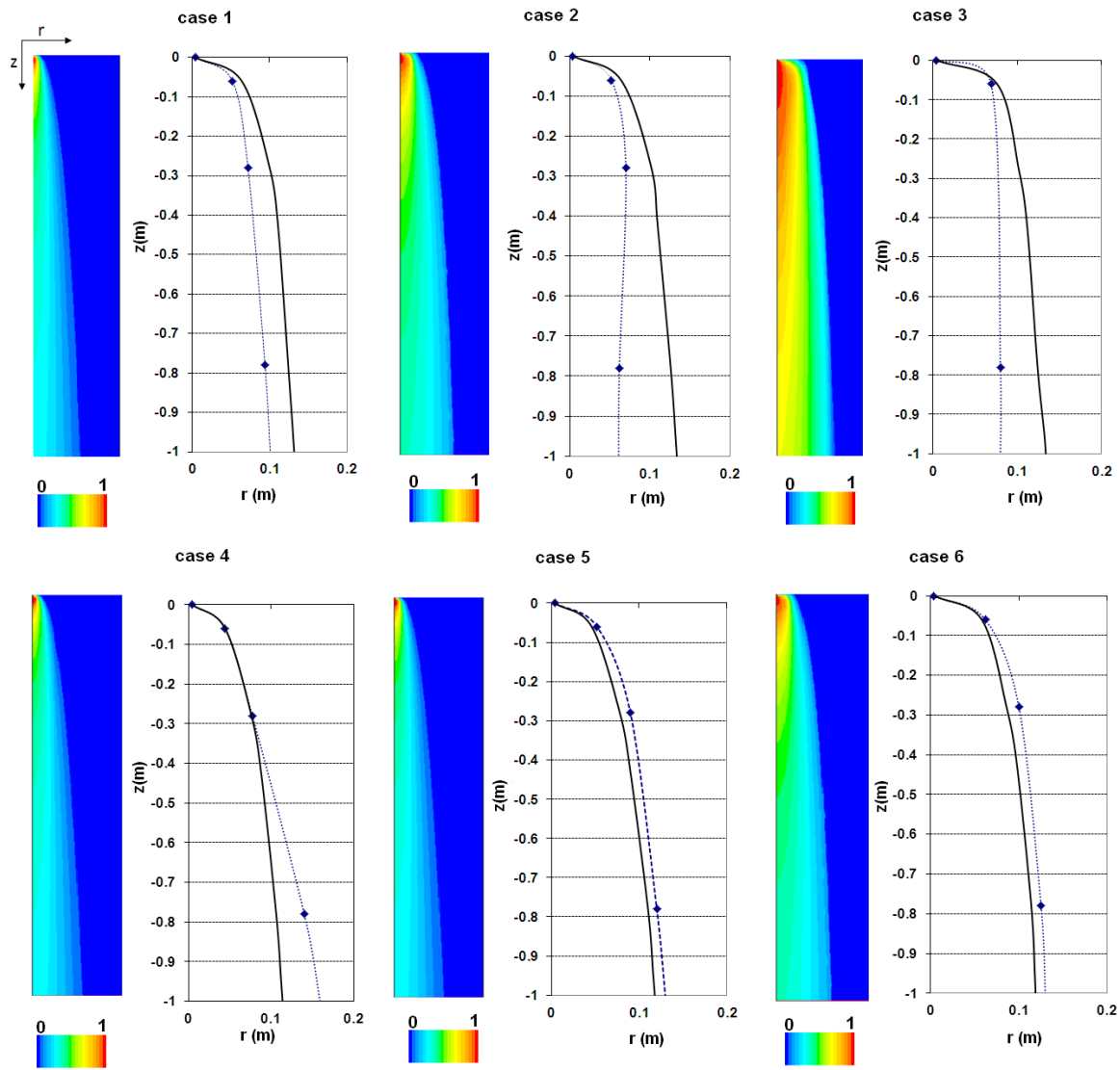

**Figure 3 : Liquid saturation fields and liquid jet contours in simulation cases described in Table 1. Discrete points: experiments; continuous curves: CFD simulations.**
**case 1/** $Q_L$=128 l/h, $Q_G$=45 m$^3$/h, $d_p$=1.99mm; **case 2/** $Q_L$=128 l/h, $Q_G$=90 m$^3$/h, $d_p$=1.99mm; **case 3/** $Q_L$=453 l/h, $Q_G$=45 m$^3$/h, $d_p$=1.99mm; **case 4/** $Q_L$=128 l/h, $Q_G$=45 m$^3$/h, $d_p$=6mm; **case 5/** $Q_L$=128 l/h, $Q_G$=90 m$^3$/h, $d_p$=6mm; **case 6/** $Q_L$=453 l/h, $Q_G$=45 m$^3$/h, $d_p$=6mm

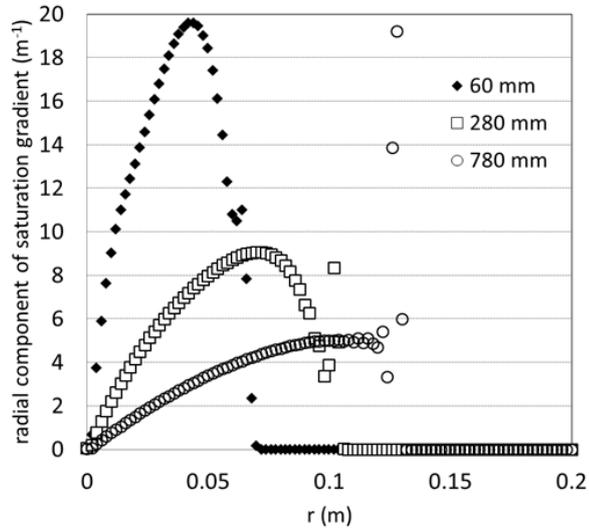

**Figure 4 : profile of radial coordinate of liquid saturation gradient at different longitudinal bed positions for $Q_L$=128 l/h, $Q_G$=45m³/h, and $d_p$=1.99mm**

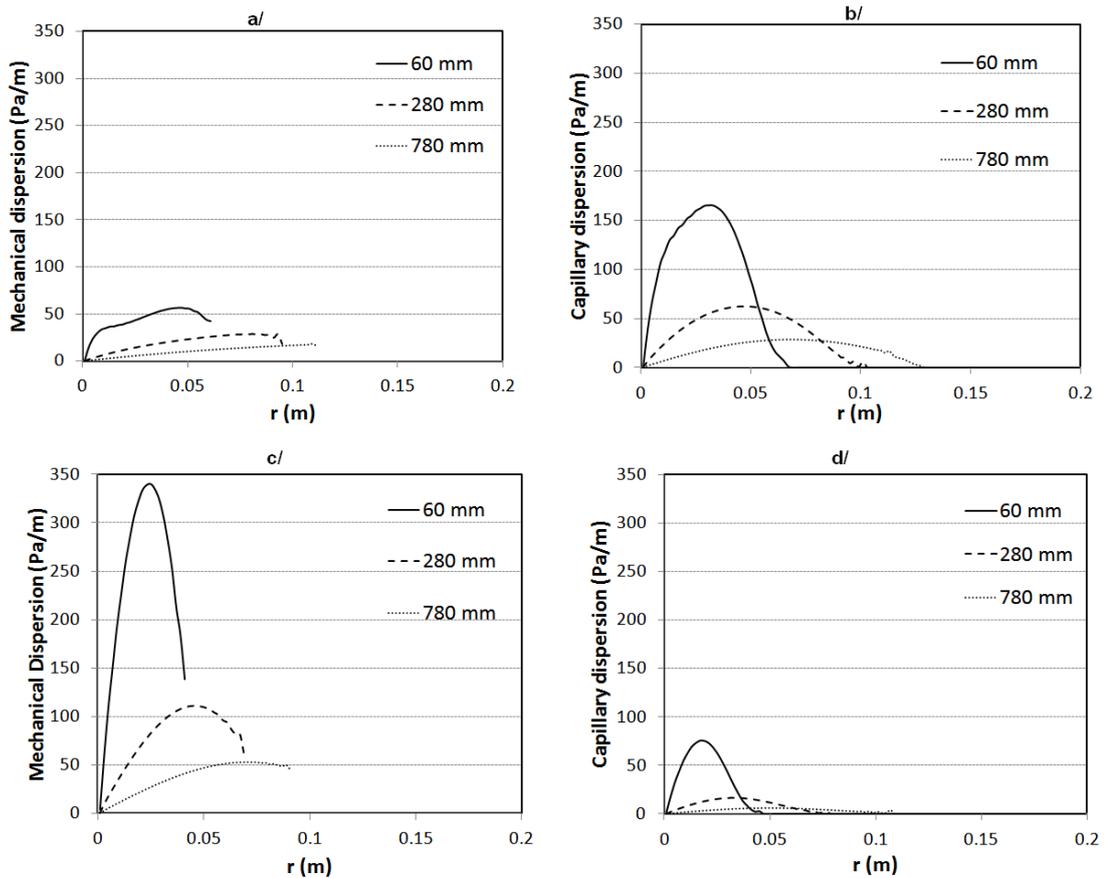

**Figure 5 : Evolution of radial component of mechanical and capillary dispersion forces for $Q_L$=128 l/h and $Q_G$=45m³/h, (a) and (b): $d_p$= 1.99 mm, (c) and (d): $d_p$= 6 mm**

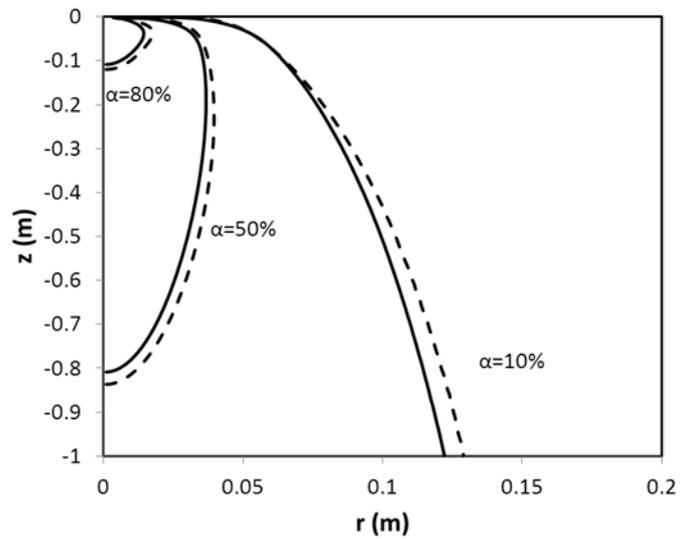

**Figure 6 : Liquid iso-saturation contours for $d_p$=1.99 mm and $Q_L$= 453 l/h, discontinuous line: Qg= 45 m3/h, continuous line: Qg= 90 m3/h**

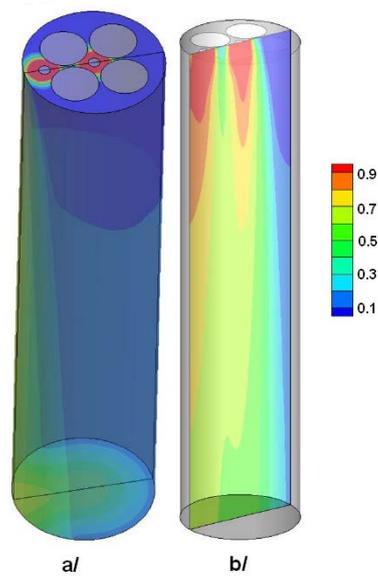

**Figure 7 : contours of liquid saturation obtained by simulation in the case of Marcandelli et al., (2000) experiment. a/ on the entire domain, b/ on the symmetry plane**

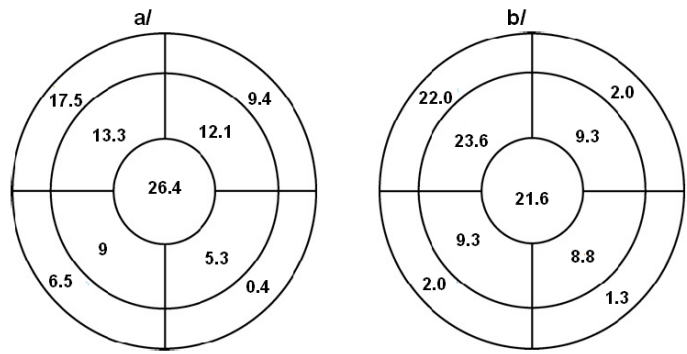

**Figure 8 : liquid distribution at bottom of the trickle-bed in terms of % of total liquid flow rate. a/ experimental results (Marcandelli et al., 2000), b/simulation results**